\documentclass[aps,prb,reprint,groupedaddress,floatfix]{revtex4-1}
\usepackage{epstopdf}
\usepackage[pdftex]{graphicx} 

\usepackage{amssymb} 
\usepackage{amsmath}
\usepackage{float}
\usepackage[table]{xcolor}

\let\baraccent=\= 
\renewcommand{\=}[1]{\stackrel{#1}{=}} 

\usepackage[normalem]{ulem}

\begin{document}

\title{ Graphene Oxide Photoreduction Recovers Graphene Hot Electron Cooling Dynamics}
\author{Alden N. Bradley$^{1*}$, Spencer G. Thorp$^{1*}$, Gina Mayonado$^{1}$, Edward Elliott$^{2}$,  Matt W. Graham$^{1}$}
\affiliation{$^{1.}$ Department of Physics, Oregon State University, Corvallis, OR, 97331, USA}
\affiliation{$^{2.}$Voxtel Nano, Corvallis, OR, 97330, USA}

\begin{abstract}

Reduced graphene oxide (rGO) is a bulk-processable quasi-amorphous 2D material with broad spectral coverage and fast electronic response. rGO sheets are suspended in a polymer matrix and sequentially photoreduced while measuring the evolving optical spectra and ultrafast electron relaxation dynamics.  Photoreduced rGO yields optical absorption spectra that fit with the same Fano lineshape parameters as monolayer graphene. With increasing photoreduction time, rGO transient absorption kinetics accelerate monotonically, reaching an optimal point that matches the hot electron cooling in graphene.  All stages of rGO ultrafast kinetics are simulated with a hot-electron cooling model mediated by disorder-assisted supercollisions. While the rGO room temperature 0.31  ps$^{-1}$ electronic cooling rate matches monolayer graphene, subsequent photoreduction can rapidly increase the rate by 10-12$\times$. Such accelerated supercollision rates imply a reduced mean-free scattering length caused by photoionized point-defects on the rGO sp$^2$ sub-lattice. For visible range excitations of rGO, photoreduction shows three increasing spectral peaks that match graphene quantum dot (GQD) transitions, while a broad peak from oxygenated defect edge states shrinks.  These three confined GQD states donate their hot carriers to the graphene sub-lattice with a 0.17 ps rise-time that accelerates with photoreduction.  Collectively, many desirable photophysical properties of 2D graphene are replicated through selectively reducing rGO scaffolded within a 3D bulk polymeric network.  \\
$^*$ co-authors contributed equally.

\end{abstract}


\maketitle

\section{Introduction}
Graphene oxides (GO) are  a widely-used substitute for graphene’s remarkable mechanical properties, but its highly amorphous lattice lacks desirable electronic properties such as high conductivity, fast photoresponse and broad spectral coverage.   When  GO is incorporated in certain polymeric networks, we show systematic photoreduction makes it more graphene-like while maintaining pristine optical-quality films.  GO\ has oxygenated functional groups attached to the 2D carbon lattice via out-of-plane bonds that prevent GO\ sheets from aggregating in solution phase.\cite{Yang2018,Mkhoyan2009}   GO can be made more graphene-like by chemical or photothermal reduction to make reduced graphene oxide (rGO).  Conventionally, these graphene-like rGO layers aggregate and scatter light strongly,  making their optical properties hard to compare against monolayer(ml) graphene. Using systematic reduction of isolated GO-in polymer composites, we show the emergence of spectral lineshapes and extract ultrafast hot-electron cooling dynamics that are closely analogous to that of ml-graphene.  

GO is often used as a bulk-processable substitute for graphene for wide-ranging applications, including electronic sensing, plasmonics, and desalination.\cite{Mueller2010, Lemme2011,Bonaccorso2010,yan2012,Blake2008,Fong2012,Wu2021a} The large presence of oxygen in GO introduces an effective band gap (Fig. 1a inset), with a tunable energy determined by the carbon-to-oxygen ratio.  Previous theoretical and experimental studies suggest bandgaps ranging from $\sim$0.6-3.1 eV for GO that can vanish nearly completely as GO is reduced.\cite{Velasco-Soto2015} GO samples reduced via pulsed Xe arc lamps effectively remove hydroxyl, epoxy, and carboxyl groups to increase the size of graphene-like $sp^2$ regions.  The amount of photoreduction changes the ratio of the  oxygenated-$sp^3$  to conjugated-$sp^2$ sub-lattice regions.\cite{Ji2012,Yang2013,Shi2014}    Very selective growths and controlled reduction are required to realize desired optoelectronic applications for GO that have included broadband optical nonlinearity\cite{Liu2009,Liu2011a}, tunable photoluminescence, \cite{Xin2012} and resonant energy transfer.\cite{Boukhoubza2019} 

 With widely-varying ratios of oxygen and carbon, the highly inhomogeneous and amorphous nature of GO and rGO lattice make a direct comparison to ml-graphene difficult. In rGO, individual $sp^2$ graphene-like sub-lattice regions often become surrounded by $sp^3$ oxidized domains, forming molecular-like confined regions often called graphene quantum dots (GQDs) or graphene nanoclusters.  While the composition of rGO varies greatly,  it can roughly be decomposed into three types of sub-lattice illustrated in Fig. 1b: (1) extended $sp^2$ hybridized regions, (2) confined $sp^2$ lattice nanoclusters or GQDs, and (3)  oxidized or $sp^3$ regions.  Zhang\textit{ et. al}  performed transient absorption on rGO in solution and found that the carbon ($sp^2$) and oxidized domains ($sp^3$) could be treated independently.\cite{Zhang2013,Wu2021} Photoexcited carriers in the spatially-confined $sp^2$ GQDs produce Frenkel excitons with energies tunable with the size of the GQD conjugation network.\cite{Kozak2016,Sk2014} The local oxygenated functional groups at domain edges also create many optically active defect states within the lattice that are seen in photoluminescence studies.\cite{Roy2017,suhaimins2022,Zhu2017}

While some of the mechanical and chemical properties of GO-based materials are analogous to graphene, the conditions necessary to replicate graphene-like electronic behavior in rGO are less clear. Past studies have compared the transient absorption (TA) response of GO and rGO prepared by chemical reduction in solution \cite{Kaniyankandy2011} and thin films. \cite
{Zhu2017,sebastian2022} This study concerns the optical properties of GO and rGO embedded in a transparent polymer film over six controlled degrees of photoreduction. The TA relaxation resolves how the ultrafast hot electron cooling rate is modified at each stage of photoreduction using tunable probe energies ranging from 1.2 to 2.3 eV.  While the hot electron cooling in graphene is typically modeled with two rates associated with optical phonon scattering and disorder-assisted relaxation processes,\cite{Strait2011,Wang2010,Graham2013} In addition to graphene-like relaxation,  prior rGO studies are dominated by a long, 10-200 ps relaxation component previously ascribed to electron trapping at defect sites.\cite{Singh2021}  

The results obtained from the succession photoreduction of GO are modeled with first-principle models of absorption lineshapes and hot-electron cooling applied previously to graphene. In Section IV.A, the evolution of the absorption lineshape with photoreduction is modeled by competing contributions from graphene-like Fano lineshape and GO-oxide-related absorption. Then Section IV.B applies a hot electron supercollision model to determine at what stage of photoreduction rGO most closely matches the dynamics of ml-graphene.  Over most visible and UV excitation energies, Section IV.C shows the GO-sub-lattice and graphene quantum-dot states dominate both the photoluminescence and ultrafast response. Lastly, we resolve how photoreduction of GO impacts the ultrafast rate of acceptor-donor electron transfer from the photoexcited GQDs to graphene acceptor states.

\begin{figure}
   \begin{center}
  \begin{tabular}{c}
   \includegraphics[height=16.5cm]{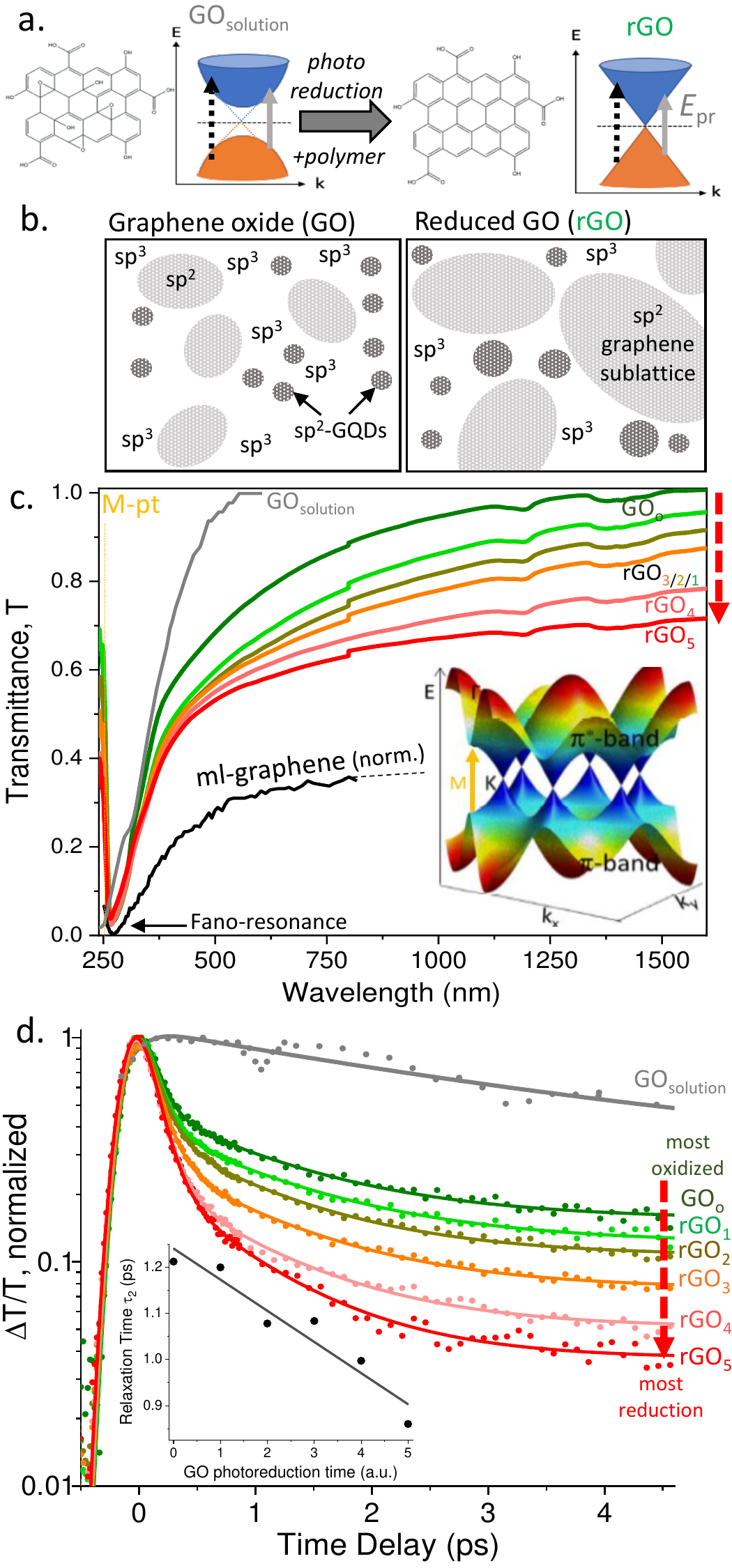}%
  \end{tabular}
   \end{center}
   \caption{\textbf{(a) }  Comparison of GO  vs. rGO  band with chemical structures. \textbf{(b)} Illustration of the three prominent sub-lattices types within the rGO structure (sp$^2$, sp$^2$ graphene quantum dot (GQD) and oxygenated sp$^3$-lattice). \textbf{(c)}  Linear and transient absorption spectra  are measured at five stages of the photoreduction. With increasing photoreduction, NIR transmittance  decreases to  more closely approximate the (renormalized)  CVD ml-graphene  transmittance curve.  Conversely,  as grown GO in solution  (gray line) has a prominent $\pi-\pi^{*}$ bandgap. (\textit{inset}) Graphene band structure highlighting the M-saddle point transition. \textbf{(d)} Corresponding transient transmittance kinetics at $E_{probe}$=1.8 eV show carrier relaxation accelerates with reduction. (\textit{inset}) The $\tau_2$ lifetime increases linearly with photoreduction.
} 
    {\label{fig1} 
}
   \end{figure}

 \section{Experimental Methods}

The GO and rGO polymer samples were fabricated using commercially available chemically exfoliated graphene oxide sheets (Graphenea) containing $\sim$53\% carbon and $\sim$44\% oxygen. The sheets are dispersed in a N, N-dimethylacrylamide (DMAA) polymer with added PMMA sites to scaffold the GO and minimize aggregation. The mixture is cured between two 1 mm thick glass slides, resulting in a sample thickness of 220 microns. The sample is then photo-reduced via a pulsed Xenon arc lamp at a 1 Hz repetition rate. This low frequency was chosen to prevent gas bubbles from forming during the reduction process. Absorbance is measured via Cary IR-UV-Vis spectrometer. Both excitation and emission photoluminescence are detected with a commercial fluorimeter (Horiba NanoLog).

Both degenerate and non-degenerate pump-probe experiments are conducted with 140 fs pulses from a Ti:sapphire lasers (Coherent Chameleon) and Optical Parametric Oscillators (APE OPO Compact). An optical parametric amplifier is used to tune the output wavelength. The beam is split into two parts, a strong pump and a weaker probe power beam with a ratio of $\sim$10:1. The intensity of the pump beam is modulated using an acousto-optic modulator (AOM, Crystal Tech) at 500 kHz. The polarization of the pump and probe beam is linear and set parallel to each other. For the non-degenerate experiment, the pump beam is frequency doubled by a second harmonic generation unit (OPE SHG) prior to modulation. Alternatively, a white-light  supercontinuum is generated to provide a broadly tunable probe.  Both beams are focused onto the sample by a single lens. The probe beam waist at the sample is approximately 80 microns.   The transmitted probe-beam is detected by photodiode lock-in amplification (Zurich Instruments, HFLI and  MFLI) at 500 kHz modulation. 

To compare the rGO polymer physics to ml-graphene, similar measurements to the above were carried out using an ultrafast transient absorption (TA) microscopy setup with a 1 $\mu$m spot size.  The ml-graphene was prepared by chemical vapor deposition (CVD) and wet-transferred to a thin silicon nitride grid. The above non-degenerate pump-probe scheme was used in a collinear geometry coupled to a 4\textit{f}-confocal scanning microscope (Olympus BX51W).  The absorption spectra of ml-graphene are taken on the same microscope by coupling in a tunable Xe-arc illumination source and detecting the full plane images on a camera (EMCCD, PI-ProEM) camera after background renormalization.

 \section{Results}



Spanning the UV to near-IR regions, Fig. 1c plots the absolute linear transmission of six graphene oxide (GO) samples in a polymer composite with increasing photothermal reduction times labeled from rGO$_{1}$ to rGO$_5$. Additionally plotted on a renormalized scale, we overlay the linear absorption spectra  of both pristine monolayer (ml) graphene (\textit{black line}),  and the starting as-grown commercial GO solution (gray line, GO$_{solution}$). The GO solution has a clear bandgap, peaking at the molecular $\pi-\pi^*$ transition.  Conversely, ml-graphene gives an expected  Fano resonance lineshape  peaked at 265 nm, red-shifted from the M-saddle-point transition labeled  in  Fig. 1c (\textit{inset}).\cite{Santoso2011} The rGO$_o$ curve in Fig. 1c is the `as-grown'  GO after incorporation into a hybrid polyacrylic and PMMA polymer matrix described in the methods. The absolute absorbance increases monotonically with GO photothermal reduction time over the NIR and IR regions plotted (from 0.35 eV to 1.5 eV).   Photoreduction of GO leads to a spectral lineshape that absorbs light more analogously to CVD monolayer graphene plotted in Fig. 1c. 

In the solution phase and most polymers,  GO\  aggregates as it is reduced, resulting in colloidal mixtures that strongly scatter light.   GO\ is incorporated in a polymer-sphere matrix scaffold that makes systematic photoreduction possible while maintaining pristine optical quality films.  Thus, we are able to  compare the absorption lineshapes, photoluminescence, and ultrafast hot electron cooling rates over a wide range of  photoreduction. Interestingly,  the more heavily reduced graphene oxide samples in Fig. 1c have a transmittance  lineshape and slope similar to ml-graphene throughout the near-infrared (NIR) regions.    In the supplementary Fig. S2, this absorption spectrum is extended out past 3 $\mu$m to the IR-region where the strong similarity to graphene absorption is maintained. 

Figure 1d plots the normalized transient transmission ($\Delta T/T$, semi-log scale) kinetics of sequentially photoreduced GO/rGO samples acquired with a 1.8 eV degenerate pump and probe configuration.  As the degree of reduction increases, the kinetic relaxation rate accelerates. The data shown in both Figs. 1 and 2 fits (\textit{solid lines}) to a least-squares algorithm requiring three-exponents ($\tau_1$, $\tau_2$, and $\tau_3$) with pulse deconvolution for the 155 fs laser autocorrelation response.   After GO is incorporated and stabilized in the polymer matrix, the relaxation dynamics accelerate monotonically with photoreduction time.   In stark contrast, the as-grown  solution of GO (gray line in Fig. 1d) has much longer TA relaxation dynamics at all timescales, bearing little resemblance to faster graphene. 

At a 1.8 eV visible probe energy, the  GO polymer composite that received no reduction (highest oxygen content) has the longest TA relaxation kinetics with its $\tau_3$ component comprising 21\% of total decay amplitude.  The inset of Fig. 1d shows the $\tau_2$ lifetimes  all decrease  linearly from $\sim$1.2 to 0.9 ps with increasing lamp photoreduction time.   All samples have a characteristic $\tau_2$ time similar to graphene's characteristic $\sim$1 ps decay expected for 1.8 eV probe, suggesting all five samples exhibit graphene-like hot-electron cooling dynamics.   By analogy with monolayer graphene, the $\tau_1$ would be associated with relaxation by optical phonons, and $\tau_2$ with disorder-assisted hot electron cooling.\cite{Graham2013}  The fitting parameter for the fast and long decays are constant at $\tau_1=0.15$ ps and $\tau_3=66$ ps,  and all parameters are shown in Fig. 2c-d.

 Figure 2 plots how the kinetic relaxation rates depend on the selected probe energy ($E_{pr}$). Comparing  Fig. 2a at  $E_{pr}$=1.3 eV to Fig. 1d at 1.8 eV, a similar pattern with photoreduction emerges.  However, the longest component, $\tau_3$ is negligible for all five cases of photothermal reduction rGO$_{1-5}$. In Fig. 2d the slower $\tau_2$ lifetime decreases linearly from 2.5 ps to 1 ps with increasing photoreduction time.  $\tau_1$ varies the least with photoreduction.  Interestingly, the most reduced samples relax even faster compared to monolayer CVD-grown graphene (\textit{\textit{black dashed line}}).  Figures 2a show  fits  to a  triexponential decay curve showing lifetimes of $\sim$0.4 ps, 1-2.5 ps, and $>$30 ps for $\tau_1$, $\tau_2$ and $\tau_3$ respectively. 
 
   \begin{figure*}[hbt]
   \begin{center}
   \begin{tabular}{c}
   \includegraphics[height=9cm]{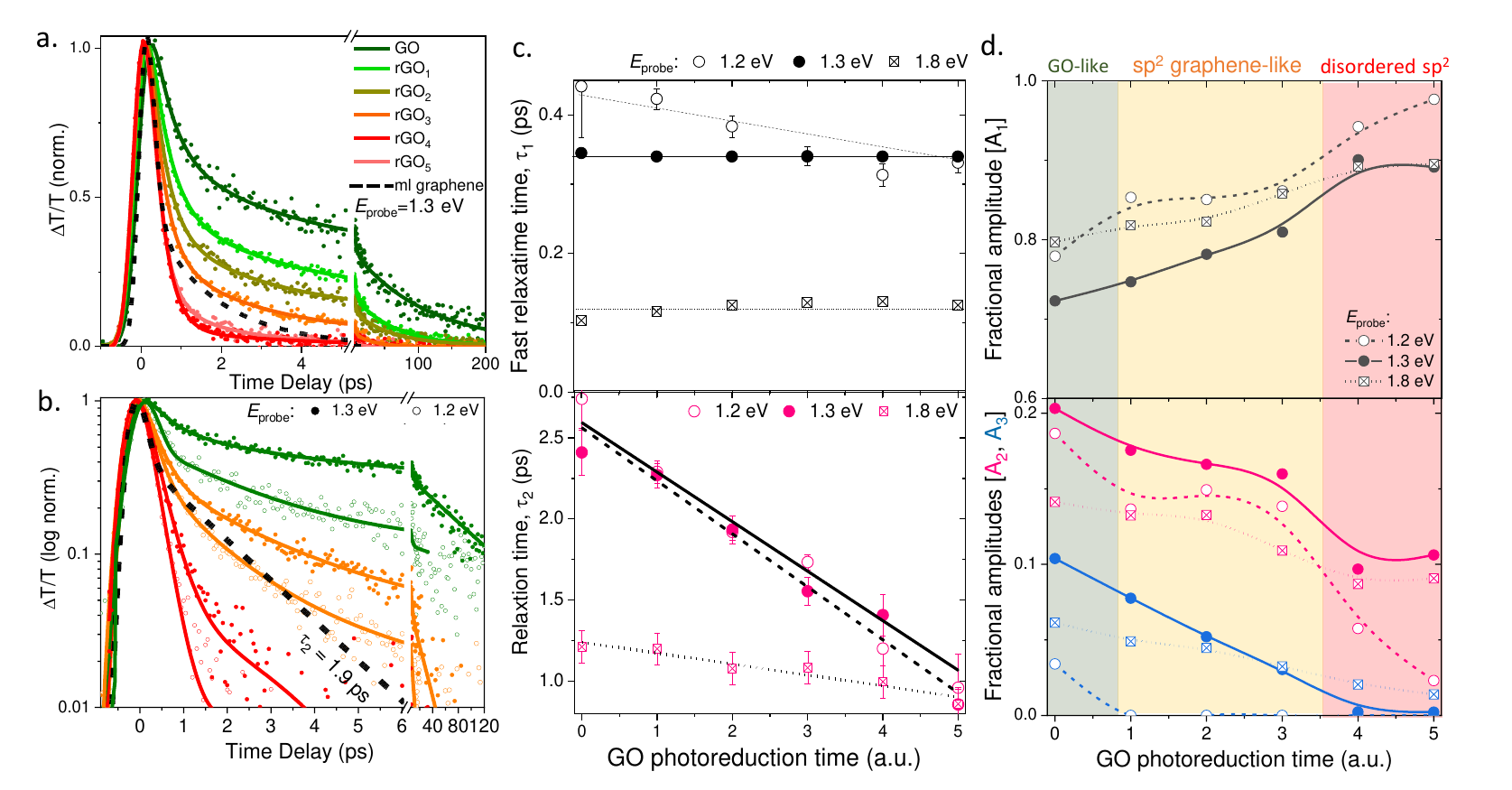}
   \end{tabular}
   \end{center}
   \caption{\textbf{(a)}  $\Delta T/T$ relaxation kinetics at $E_{probe}=$ 1.3 eV accelerate with sequential GO photoreduction.  Fits show two exponential lifetimes, with only the most oxidized samples requiring a third lifetime of $\tau_3=$ 61$\pm$2 ps.\textbf{(b)} The $\Delta T/T$ kinetics for $E_{probe}=$ 1.2 eV  (open circles) relax faster than at 1.3 eV (\textit{closed circles}). The rGO$_3$ photoreduction stage most closely approximates the ml-graphene interband relaxation kinetics shown (\textit{dashed line}).   \textbf{(c)} For each probe energy, the  $\tau_1$ lifetimes (top) are roughly constant, whereas the  $\tau_2$ lifetime (bottom) decrease linearly $\sim2.5 \times$ during with photoreduction to become even faster than ml-graphene. \textbf{(d)}   Amplitudes ($A_{1/2/3}$) of each lifetime component suggest a composition change with increasing amplitude from sp$^2$ sub-lattice dynamics.  The  smallest $A_3$ (blue) amplitude quickly decreases to zero as GO is reduced. } 
    {\label{fig2} 
}
   \end{figure*}

  Regardless of the incident TA probe energy (1.2 to 1.8 eV), rGO samples relaxed progressively faster as the photoreduction time increased.  Figure 2b  shows that TA dynamics of GO, rGO$_3$, and rGO$_5$ are slower at $E_{pr}$= 1.3 eV (closed circles, 2.6 eV pump)  than the  $E_{pr}$= 1.2 eV  (open circles, 2.2 eV pump) probe energy window.  Interestingly, the most reduced sample, rGO$_5$, always decays more quickly than ml-graphene. This faster decay relative to graphene suggests that the photothermal reduction is ultimately damaging the sp$^2$ graphene-sub-lattice by causing increased disorder and defect sites.  This symmetry-breaking results in low energy disorder states that have been previously observed in conjugated carbon systems .\cite{Skomski2014, Bhaumik2017} This is further supported by the  qualitative increase in lattice defect states that is evident by increased emission in  IR region of the PL spectra (see supplemental Fig. S2).
  
  Figures 2c-d contain the results of our  exponential fitting lines  shown in Fig. 1d and 2a-b (\textit{solid lines}). The top panel shows the amplitude of the fast time component ($\sim$0.4 ps) at 1.2 eV, 1.3 eV, and 1.8 eV, which accelerates only moderately  as the GO samples are reduced. The middle panel shows the amplitude of the second ($\tau_2 \approx 1-$2.5 ps, pink) and third ($\tau_3 > $30 ps, blue) time components, which both decrease with reduction.  Importantly,  the slow time $\tau_3$ component goes to zero in the limit of heavy reduction and closely resembles the ml-graphene relaxation. The bottom panel of Fig. 2c shows the  $\tau_2$ relaxation time of GO  decreases roughly linearly with photoreduction time.   At all probe energies, the  $\tau_2$ relaxation time decreases with reduction, with rGO$_{3,4,5}$ having lifetimes shorter than that of CVD graphene under the same optical conditions.  The CVD ml-graphene (dashed line in Fig. 1-2) was fit to a $\tau_2=$1.9 ps at 1.2 eV and 1.1 ps at 1.8 probe energies respectively.    

 In most heavily oxygenated rGO samples, the longest $\tau_3 \sim 61$ ps component comprises up to 16\%  of the total decay amplitude.   Such samples contain many functional groups, however, the large band gap of the fully oxided regions is well outside the spectral range of both pump and probe laser energies.  Instead, graphene quantum dots (GQD) create gapped sp$^2$ molecule-like regions with size-tunable bandgaps that are resonant with our probe beam.\cite{Roy2017} For rGO$_{3,4,5}$ samples,  Fig. 2d shows that the $\tau_3$ time-component is zero for $E_{probe}<1.3$ eV, suggesting only  graphene-like sp$^2$ sublattice regions are relevant to the electronic dynamics throughout this near-infrared probe region.  

\section{Analysis and Discussion}

\subsection{rGO  Fano Lineshape  Absorption Analysis  }

\begin{figure}[h]
   \begin{center}
   \begin{tabular}{c}
   \includegraphics[height=6cm]{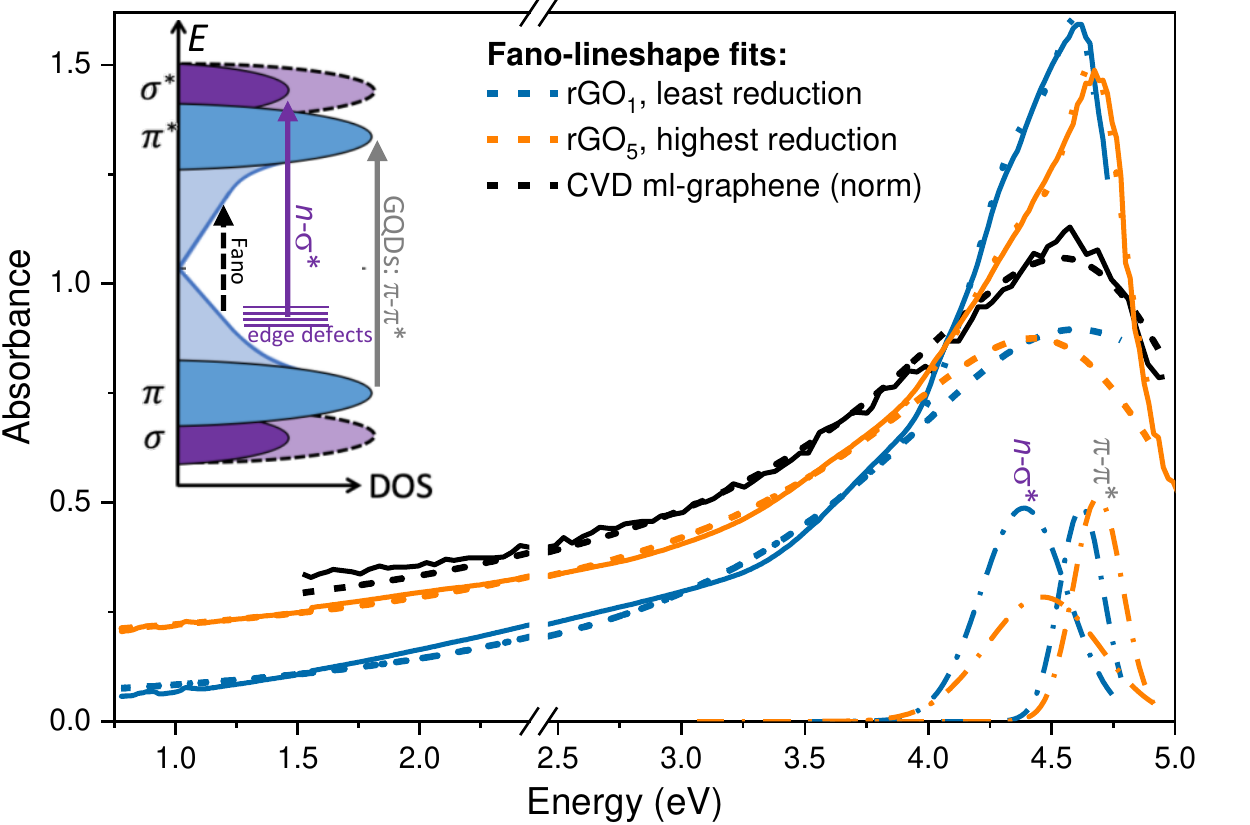}
   \end{tabular}
   \end{center}
   \caption{ Under each linear absorption spectra (\textit{solid lines}) the deconvolved Fano resonance  lineshape fit is plotted in \textit{dashed lines}. Unlike pristine ml-graphene (\textit{black}), the two rGO samples plotted also require two convolved Gaussians (\textit{dash-dot}) suggesting molecular-like transition labeled $\pi$ to $\pi^*$ and edge defect transitions, $n$ to $\sigma^*$  (see inset).  The resulting Fano-Gaussian convolved fits (\textit{dotted lines}) show the graphene sub-lattice Fano-parameter, $q$ increases with photoreduction consistent with more lattice disorder. }
  

    {\label{fig3} }
   \end{figure} 
   
  The  transmission spectra in Fig. 1a, Fig. S2 and fitted absorption spectra in Fig. 3 all show lineshapes similar to ml-graphene  throughout the NIR and IR spectral regions from $\sim$0.4 to 3.5 eV. The absorption maxima of both ml-graphene and rGO  in Fig. 3 (black line) deviate  from the tight-binding model prediction of the graphene van Hove singularity M-point resonance at $\sim$5.1 eV.\cite{CastroNeto2009}  Instead,  the graphene absorption is best fit by a Fano lineshape with a renormalized peak resonance energy, $E_r$ that is red-shifted from the M-point by energy by $\cong 0.3-0.4$ eeV.\cite{Yang2009,Mak2010} The asymmetric Fano lineshape accounts for the ratio of interference between the discrete (M-point) and continuum  transition probabilities through the dimensionless Fano parameter, $q$.\cite{Yang2009}  Thus, the tight binding model of the graphene absorption spectrum in Fig. 3 is renormalized for effective electron-hole interaction effects by fitting to the below asymmetric Fano lineshape,
\begin{equation}
A_{Fano}(E)=A\left[\frac{\left(\frac{2}{\gamma}(E-E_r)+q\right)^2}{1+\left(\frac{2}{\gamma}(E-E_r\right)^2}\right]
\end{equation}
where $\gamma$ is the Lorentizian homogeneous linewidth and \textit{A} is the amplitude scaling constant.  Fig. 3 plots a hyperspectral measurement of CVD ml-graphene (black line) with its corresponding Fano lineshape fit (dashed line), given by equation 1 above. Table I gives the resulting Fano parameters and show excellent agreement of this work graphene values with the established literature values.\cite{Yang2009,Mak2008}
This provides an essential calibration base to quantitatively compare against the lineshape fit of rGO absorption spectra.

Figure 3 shows good agreement between the absorption spectra of rGO$_1$ and rGO$_5$, and the asymmetric Fano resonance after it is convolved with 
two Gaussians peaks at energies corresponding to the absorption of the $n-\sigma ^*$ and $\pi-\pi^*$ transitions.  This fitting analysis suggests that the absorption spectrum in rGO can be understood to contain a Fano resonance similar to that of CVD ml-graphene.  The molecular-like  $\pi-\pi^*$ transitions are illustrated in Fig. 3 (\textit{inset}), and show graphene quantum dot (GQD) states also contribute to the spectral weight and are centered near 4.6  eV.\cite{Chien2012} At 4.3 eV, rGO also contains sub-gap defect states between the $\pi$ and $\pi^*$ states, which results from previously reported local oxygen-based disorder that creates edge defect state ($n$) to $\sigma^*$ transitions.\cite{Sehrawat2018,Yoon2016,Liang2015,Sutar2012, Mkhoyan2009} Due to the heterogeneous oxygen coverage, these local disorder edge states have a much broader absorption FWHM.  As rGO$_1$ is further reduced, we observe in Fig. 3 that the peak area of the $n-\sigma^*$ Gaussian decreases as oxygen is removed, resulting  in fewer edge states.    Both our most oxidized samples (GO$_{o}$ and GO$_{solution}$) did not fit well to a Fano lineshape, suggesting only rGO samples have a graphene-like absorption lineshape in the IR\ and NIR regions.

Table I contains a summary of the Fano fitting parameters, showing good agreement between the literature\cite{Mak2008,Yang2009} and our results for monolayer graphene and rGO$_5$. rGO$_5$ contains a large absorption from the linear dispersion near the K and K' points, where excited carriers couple strongly to the continuum, similar to monolayer graphene. For rGO$_1$, the Fano parameter \emph{q} decreases significantly from monolayer graphene, suggesting electron-hole interaction effects are increasingly screened for transitions near the van Hove singularity.  For GO and lightly reduced rGO, Table I shows the Fano parameter is many times larger than highly reduced samples and monolayer graphene.  This suggests the many edges states in more oxidized graphene couple strongly to continuum-like states.   
\begin{table}[t]
\begin{tabular}{ p{3.9cm} p{1.8cm} p{1.5cm} p{1.3cm} }
\hline
 Sample & $E_r$ (eV) & $\gamma$  (eV) & \emph{q} \\

\hline
ml-graphene [CVD]  & 4.80 & 1.69  & -3.2\\
ml-graphene [exfoliated]\cite{Yang2009} & 4.73 & 1.30 & -3.3\\
rGO$_5$  [highly reduced]    & 4.69 & 1.68 & -3.2 \\
rGO$_1$ [barely reduced]  & 4.62 & 2.16 & -50 \\
\hline
\end{tabular}
\caption{Fano fitting parameters for data in Fig. 3 (\textit{dashed lines}) show good agreement with our monolayer graphene data established literature values.\cite{Yang2009,Mak2010} The Fano parameter \textit{q} of rGO$_5$ best matches ml-graphene.  Two convolved Gaussian for GQD $\pi-\pi^*$ and edge state defects are also required. }
\label{Table 1}
\end{table}

The inset of Fig. 3 shows a qualitative depiction of how the density of states changes from GO  to rGO. As the samples are reduced, they contain larger area regions of non-interrupted $sp^2$ carbon, leading to a more graphene-like distribution of continuum states, resulting in a better Fano lineshape fit. The two convolved Gaussians  show the effect of reduction on the absorption spectra, with the amplitude of the  $n-\sigma ^*$ transition decreasing significantly, suggesting the removal of oxygen functional groups. We also see that the absorption peak in rGO$_1$ shifts slightly to lower energy compared to rGO$_5$. This shift has been theoretically predicted by Roy et al.\cite{Roy2017}, who used DFT to calculate the band structure of GO at varying oxygen content, finding that the addition of oxygen decreases the band gap at the M-point. However, the   underlying  Fano resonance energy  ($E_R$ in Table I) does not change with photoreduction.  The very large q Fano parameter required to fit the most oxidized rGO$_1$ samples suggests the sp$^2$ hybridized regions are not extensively delocalized and retain a molecular-like character.

\subsection{Hot-electron cooling rates  in reduced graphene oxide }


Figure 4 fits the hot electron cooling TA kinetics in progressively reduced GO as the TA probe energy is increased from 1.2 (top) to 1.8 eV (bottom).   Specifically, the hot-electron cooling rate ($\tau_{SC}$) is extracted.  Unlike the exponential rate $\tau_2$ from Fig. 2, $\tau_{SC}$ is analogous to the recombination rate as the electron cool near the Fermi energy, and is independent of probe energy ($E_{probe}$).  To connect the above phenomenological exponential relaxation models of GO to this first-principle hot-electron cooling model, the fits in Figure 4 models our TA relaxation kinetics using a hot electron heat dissipation rate $H=C_e(dT_e/dt)$, where $C_e$ and $T_e$ are the electronic heat capacity and temperature respectively.  The top-panel of Fig. 4a contains first-principle hot electron cooling model fits (\textit{solid lines}) to  the normalized TA kinetics of the rGO samples.    Hot electron cooling rates in rGO can be qualitatively understood by comparing to CVD ml-graphene kinetics (black dotted line). The lowest energy probe ($E_{pr}$=1.2 eV) in the top panel of Fig. 4a shows the hot electron cooling rate response of ml-graphene (dashed line) is identical to rGO$_{1}$, rGO$_{2}$ and rGO$_{3}$.  Interestingly,  rGO$_{4,5}$ dissipates heat  even  faster than CVD ml-graphene.  

   The mechanism for fast energy dissipation or hot-electron cooling in graphene has been widely debated in the past.  The optical phonon dissipation model\cite{Wang2010, Rana2011,Huang2010} evolves on the sub-ps relaxation timescale of the  $\tau_1$  component. At longer relaxation times, the disorder-mediated acoustic phonon decay pathway or supercollision (SC) hot electron cooling model are the primary factor limiting cooling of the photoexcited hot electron temperature, $T_e(t)$.\cite{Song2011a} Experimental studies demonstrate the SC-model\cite{Song2011a} successfully predicts graphene's photocurrent\cite{Graham2013}, optical \cite{graham2014} and electrical\cite{Betz2013} heating response. However, the applicability of the SC-model to more disordered lattice of GO and rGO has not been considered.   
   
To understand hot electron cooling in rGO, we apply the acoustic phonon SC-model illustrated in Fig. 4b (inset). In the  SC model, hot electron cooling near the Fermi level occurs without crystal momentum conservation. Instead, higher-energy ($\sim k_B T_e$) acoustic phonons are emitted with the momentum imbalance, $q_{recoil}$ accounted for by disorder-induced intrinsic lattice recoil.\cite{Song2011a}  This SC- hot electron is illustrated in Fig. 4b (inset), and gives in a faster hot electron cooling rate than a hot-phonon model that is  given by,\cite{Song2011a,Graham2013} 
\begin{align}
\frac{dT_e}{dt}=-\frac{H}{\alpha T_e}=-\frac{A}{\alpha}\frac{T_e^3-T_l^3}{T_e}.
\end{align}
where $A/\alpha$ is the SC rate coefficient, $T_l$ and $T_e$ are the lattice and electron temperatures, respectively. 
Solving Eq. 2, $T_e(t) \cong \frac{T_o}{1+AT_ot/\alpha}$ when $T_e(t) \gg T_l$, where $T_o$ is the initial electron temperature. Since all data shown is at $T_l=$292 K,  the transient change in $T_e(t)$ is small compared to $T_l$, or  $T_e(t)-T_l \ll T_l$ such that we can approximate Eq. 2 by expanding the leading terms to arrive at the room-temperature hot electron temperature, $T_e(t) \cong T_l+(T_o-T_l)e^{-t/\tau_{SC}}$, to get the expression $\tau_{SC}^{-1} = 3AT_l/\alpha$. \cite{Graham2013}

\begin{figure}
   \begin{center}
   \begin{tabular}{c}
   \includegraphics[height=16cm]{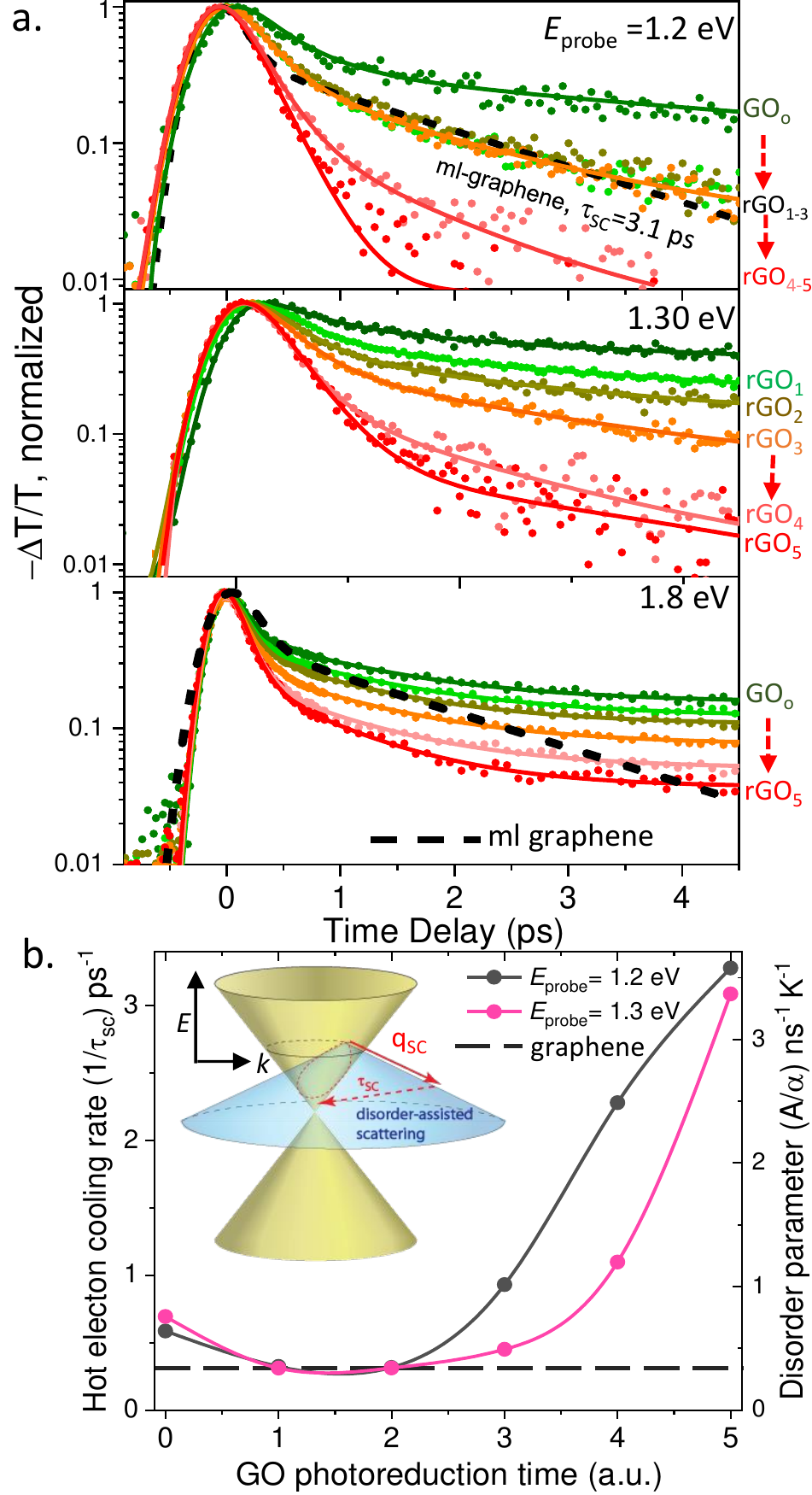}
   \end{tabular}
   \end{center}
   \caption{\textbf{(a)} TA relaxation kinetics of the six progessively reduced GO 
   GO samples compared to ml-graphene (black dashed) at 1.2, 1.3, and 1.8 eV probe energies (top to bottom). Fitted lines now incorporate the SC-hot electron cooling model of Eq. 2. \textbf{(b)} SC-model hot electron cooling rates ($\tau_{SC}^{-1}$) extracted increase sharply for longest photoreduction times. For rGO$_{1-3}$, the disorder parameter $A/\alpha$ is similar to ml-graphene rate (dashed), and expectantly is invariant to the laser probe energy of 1.2 eV (black) and 1.3 eV (pink).\cite{Graham2013}
 }
    {\label{fig5} }
   \end{figure}

The TA response is obtained using the  hot electron (or hole) temperature ($T_e$) through analytically fitting to the transient interband optical conductivity, $\Delta \sigma(E_o,t)=-e^2/4\hbar \left[ f_{e/h}(T_e(t),E_{pr})-f_{e/h}(T_l,E_{pr}) \right]$.\cite{CastroNeto2009}  The Fermi-Dirac hot-electron occupancy function, $f_{e/h}(T_e(t),E_{pr})$ at the probe energy ($E_{pr}$) equations are given in the Supplementary Materials as a change in interband optical conductivity $\Delta \sigma(t,E_{pr})$. \cite{Malard2013,CastroNeto2009}  
  In Fig. 4b the hot electron cooling rates ($\tau_{SC}^{-1}$) for rGO are extracted by fitting the data in Fig. 4a  to the analytical SC-model solution (Eq. 2), allowing for two additional exponential components ($\tau_1$ and $\tau_3$).  This fast component, $\tau_1 \cong 0.34$ ps, averages over the initial electron thermalization and optic phonon emission timescale and is discussed elsewhere.\cite{Breusing2009,tielrooij2013}  Any molecular-like $\pi-\pi^*$ transitions present are captured by $\tau_3\sim 61$ ps. 

The accelerating TA relaxation kinetics in Fig. 4a are consistent with the idea that photoreduction of GO creates more disorder and defects on the graphene sub-lattice. Figure 4b shows an increase in the rate of hot electron cooling, $\tau_{SC}^{-1}$.   Unlike the earlier exponential fits, the rate $\tau_{SC}^{-1}$  is independent of the probe energy and is the rate at which the hot-electron Fermi-Dirac distribution cools. The hot electron cooling time for the  comparison  monolayer CVD -grown graphene (dashed line in Fig. 4b) at 292 K is  3.1 ps.  $\tau_{SC}^{-1}$ increases by a factor of $\sim$6 as the samples are reduced. This suggests the xenon arc lamp used to reduce GO is a largely destructive process to underlying sp$^2$ sub-lattice. At the highest level of photoreduction, Fig. 4b suggests the increased lattice disorder destroys the desired graphene-like extended lattice by creating to many point-defects.

The $\tau_{SC}^{-1} = 3AT_l/\alpha$ expression is a direct measure of lattice disorder by the expression $\frac{A}{\alpha} \cong \frac{2}{3}\frac{\lambda}{k_F l} \frac{k_B}{\hbar}$, where the  mean free scattering path is $k_F l$.\cite{Song2011a}  The electron-phonon coupling strength can be approximated as $\lambda=\frac{D^2}{\rho s^2} \frac{2E_F}{\pi(\hbar v_F)^2}$, where both the deformation potential, $D$ and Fermi energy $E_F$ are the experimental variable that increase the hot electron cooling rate. Figure 4b shows that $\frac{A}{\alpha}\cong 0.3$ ns$^{-1}$K$^{-1}$  for rGO$_{1-3}$, which matches the monolayer CVD graphene values in literature.\cite{graham2014}  However, further photoreduction increases $\frac{A}{\alpha}$ upto 6$\times$, suggesting the graphene sub-lattice is being damaged. If the deformation potential is approximately constant, then  that $A/\alpha \propto E_F/k_Fl $, suggesting that the damage of photoreduction decreases the mean free scattering path by photoionization, which increase sp$^2$ sub-lattice defect sites. Our fitted data  in Fig. 4 confirms that acoustic phonons supercollisions (SCs) best describe the rate-limiting heat dissipation kinetics in reduced graphene oxide.  Furthermore, Fig. 4b shows how disorder from photodamage to the rGO lattice systematically increases the hot-electron cooling rate.   This controlled change in lattice disorder provides new evidence of the predominant  role of disorder-assisted SC in describing the hot-election in graphene.      

\subsection{Oxygenated sub-lattice contributions from graphene quantum dots}

Sections IV.A and B above both show the rGO sample and ml-graphene have remarkably similar lineshape and hot-electron cooling rates over optical energies that ranging  from  0.4 to 1.8 eV.  This section focuses one the differences that arise in visible and UV range where GQDs and defect-edge states are also also optically be excited.   Figure 5a plots the PL emission spectra of the least reduced, GO$_o$ and most reduced, rGO$_5$ samples after a 4.6 eV excitation.  The main asymmetric peak appears to shift from $\sim$2.4 to 2.7 eV with photoreduction. The experimental emission spectra (dots) are fit (solid lines) using 4 convolved Gaussian peaks (dotted lines). All peak energies and FWHM spectral width (except at 2.34 eV) are found to be approximately invariant to photoreduction. The peak at 2.7 eV in Fig. 5a corresponds with emission from the smallest graphene quantum dot states (labeled GQD$_1$) $\pi^*-\pi$ orbital relaxation.  At the lower energies, both peaks centered near 1.55 eV and 1.80 eV grow with photoreduction, consistent with emission from larger graphene quantum dot states labeled GQD$_2$ and GQD$_3$, respectively. We observe an increase in the emission intensity from these three $sp^2$ peaks with reduction, confirming they do not result from oxygen groups.  Conversely, the emission at 2.3 eV represents the carrier recombination in $sp^3$ oxygen ($\sigma^*-n$). The magnitude and width of this emission decrease with reduction as oxygen functional groups are removed.





\begin{figure*}[hbt]
   \begin{center}
   \begin{tabular}{c}
   \includegraphics[height=5.5cm]{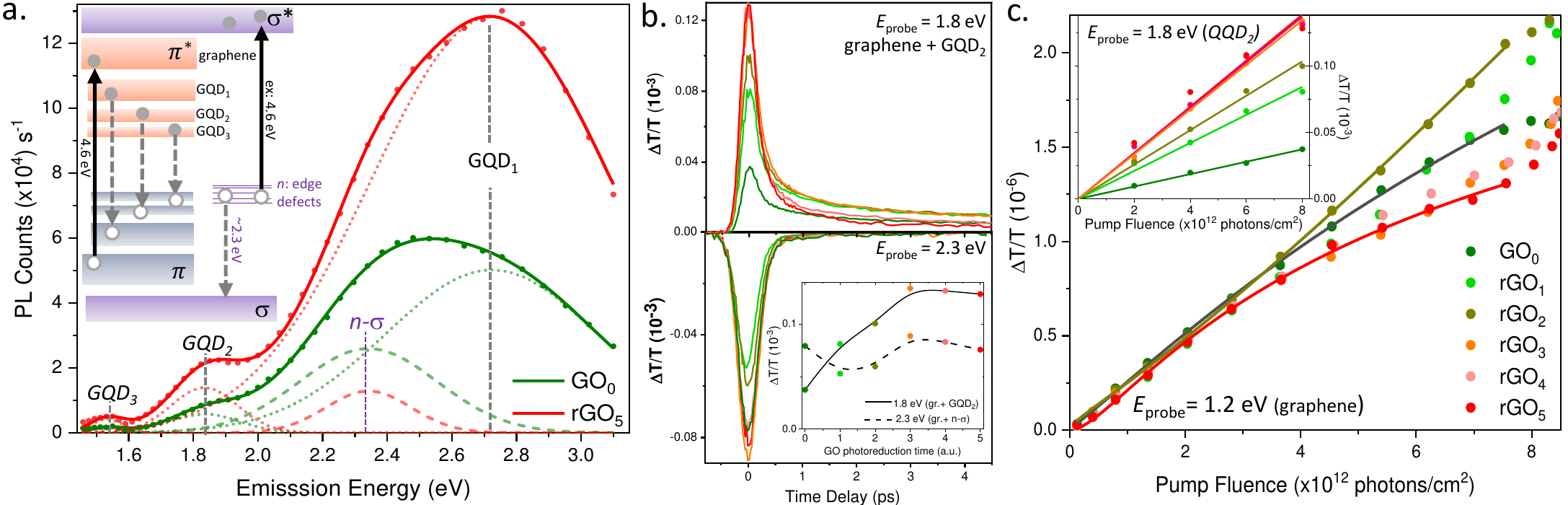}
   \end{tabular}
   \end{center}
   \caption{\textbf{(a)} The photoluminescence emission spectra of GO$_o$ (green line fit) and rGO$_5$ (red line fit) with 4 convolved  Gaussian fit (dashed lines).Photoreduction increases the PL peaks from graphene quantum dots resonances, labeled GQD$_{1-3}$. Conversely, emission from the oxygenated sub-lattice $n-\sigma$ defect edge state decreases as GO is reduced (see inset for corresponding transitions). \textbf{(b)}The degenerate TA response near the 1.8 eV GQD$_2$ resonance (top) vs. near the excited state absorption at 2.3 eV (bottom). (\textit{inset}) TA response increasing with photoreduction. \textbf{(c)} $\Delta T/T$ pump power photon fluence dependence of reduced GO samples using a 1.2 eV probe.  Fit are to the graphene SC-hot electron cooling model in Eq. 2. Over a wide range of incident photon flux, the saturable absorption susceptibility, $\Delta T/T$ is invariant to photoreduction suggesting only graphene hot electrons are probed below $\sim$1.2 eV. (\textit{inset}) Conversely at 1.8 eV the $\Delta T/T$ changes strongly, suggesting increasing GQD$_2$ states. }
    {\label{fig6} }
   \end{figure*} 
PL from GO and rGO in solution has been widely documented in the literature, showing that reduction of GO increases PL intensity at near IR wavelengths while also blue-shifting the main peak.\cite{Luo2009,Chien2012,Wang2017}  In accordance with literature, Fig. 5a shows an increase in PL intensity with reduction, at peaks centered at 1.80 eV and 1.55 eV. The PL of the oxygenated GO lattice is known to emit broadly near 2.4 eV  with locally varying oxygen content responsible for the broader FWHM.\cite{Liang2015,sebastian2022} In rGO, PL is dominated by $\pi^*-\pi$ carrier recombination in regions of confined graphene quantum dots. As the reduction process removes oxygen, formerly isolated sp$^2$ carbon atoms join together to form conjugated carbon rings, and regions that already contained large area conjugated sp$^2$ carbon structures increase in size. The observed decreasing area of the peak at 2.3 eV with photoreduction suggests this peak emission is likely due to egde states or oxygen-defects the boundaries of the sp$^3$ region. The newly formed GQD in rGO are ascribed to the increasing PL at 2.7 eV, 1.80 eV and 1.55 eV peaks. 

DFT studies by  Sk et al. \cite{Sk2014} show how the bandgap energy of a GQDs changes with respect to its size and found that GQDs about 1.3 nm in mean diameter create Frenkel exciton states near 2.7 eV, while slightly larger 2 nm GQDs emit around 1.8 eV. rGO contains an ensemble of GQDs of various sizes separated by oxygenated regions. Reduction removes oxygen, gradually increasing the GQD size, evidenced by the increased PL in rGO at 1.55 and 1.80 eV. 

Figure 5a inset contains a qualitative depiction of the bands and energy levels in GO. The optical response of graphene is determined by the $\pi$ and $\pi^*$ states, which lie between the $\sigma-\sigma^*$ gap in GO.\cite{Liang2015,Exarhos2013} Oxygen functional groups break the symmetry of the pristine graphene lattice, resulting in localized defect states that exist in the $\pi-\pi^*$ gap. Since the gap between $\sigma$ states is much larger than 2.4 eV, this emission is suggested as $n-\sigma$ transition (dashed purple arrow). In both GO and rGO, emission at 2.7 eV dominates the PL spectra, which was shown to result from $\pi$ states in isolated $sp^2$ domains (gray dashed arrow).\cite{Chien2012} Emission at lower energies comes from a broad range of GQD states and the local disorder states.

Figure 5b shows the degenerate transient absorption response of the samples at 1.8 eV and 2.5 eV, respectively. At 1.8 eV, we observe a saturable absorption signal containing a long component that slowly goes away with reduction. At 2.5 eV, we see a reverse saturable absorption response, which decays extremely quickly in all samples. A similar transition has been previously observed by Bhattacharya et. al, who saw that  a sign flip in the pump-probe response occurred near 2.3 eV.\cite{Bhattacharya2018} Since the most reduced samples have the largest reverse saturable absorption response, we can rule out excited state absorption from oxygen groups as the cause of the sign flip. We attribute this sign-change to absorption from the interband transition in graphene, which has been previously documented to exhibit a sign flip for high pump fluences at this energy.\cite{Malard2013,Gatamov2020} We do not see a change in sign when probing the oxygen states at 1.8 eV, further confirming the sp$^2$ nature of the peak labeled GQD$_2$.

Figure 5c shows the 1.2 eV probe energy  pump fluence dependence. At low pump fluences, the TA response of all samples exhibits a linear dependence on the pump fluence. Above incident photon flux of $\sim4\times10^{12}$ photon/cm$^2$,  a sublinear trend is observed that is fit to the  Eq. 2 hot electron cooling model TA response.  The nonlinear saturation effect fits to the expected nonlinear Fermi-Dirac filling factor.  Notably, the more oxidized GO$_1$ and rGO$_2$ samples have the most nearly linear behaviors, consistent with the expected smaller confined sp$^2$ sub-lattice regions. Conversely, Fig. 5c(inset) shows the pump power dependence for the differential transmission at 1.8 eV pump and probe.  GO displaying the smallest response, which increases with reduction until rGO$_3$. The response saturates for the three most reduced samples as shown in the inset of Fig. 5b. This trend matches the absorption spectra at 1.8 eV, where the absorption increases monotonically with reduction, with the exception of the $\Delta$T/T response saturating for the most reduced samples.

The pump dependence gives us insight into how the probe response changes with lattice temperature. At low pump powers, the 1.2 eV probe has the same magnitude for all samples, suggesting that even oxidized samples have large regions of graphene-like sp$^2$ hybridization. The 1.8 eV data remains linear overall pump fluences but has a large dependence on the amount of reduction. While the 1.2 eV data probes graphene-like states,  the 1.8 eV data primarily probes the confined GQD$_2$ states that lead to longer lifetimes and a strong dependence on photoreduction. The size and population of these GQD states depend heavily on the oxygen content. As shown in Figure 5c(inset), reduction increases the transient response, which suggests that reduction increases the population of sp$^2$ GQD$_2$ states that absorbs at 1.8 eV. This trend matches the increase in PL seen at 1.8 eV after photoreduction.
\begin{figure}[ht]
   \begin{center}
   \begin{tabular}{c}
   \includegraphics[width=8.5cm]{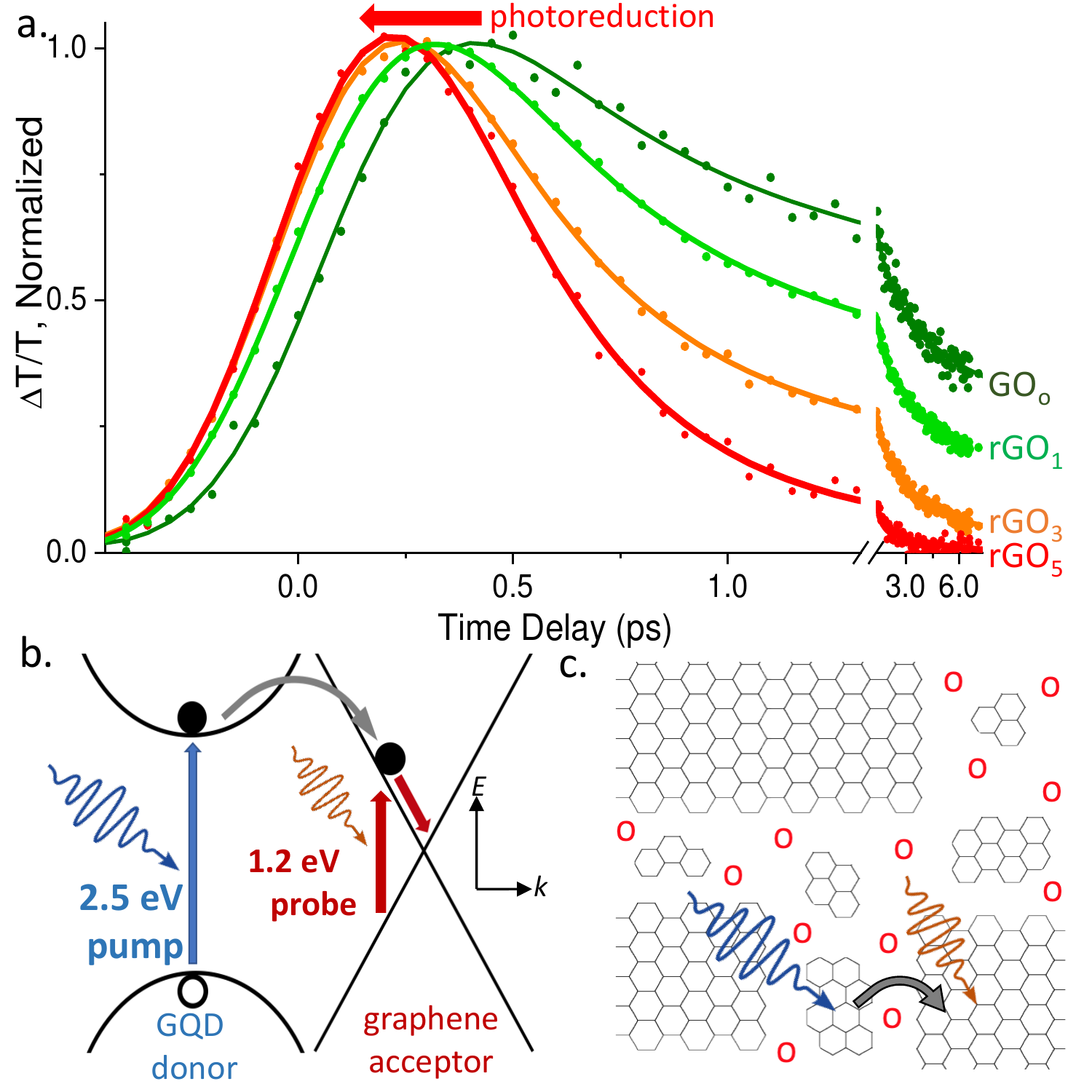}
   \end{tabular}
   \end{center}
   \caption{\textbf{(a)} Normalized transient absorption kinetics shows a 170 delayed rise for GO that systematically accelerates with successive photoreduction. \textbf{\textbf{(b}) }This rise is assigned to an acceptor-donor relationship between the 2.5 eV pump of GQD states and the 1.2 eV probe of the accepting graphene states. \textbf{(c)} Band illustration of rGO depicts charge transfer described from  confined GQDs to larger sp$^2$ graphene-like regions.}
    {\label{fig7} }
   \end{figure} 
   \subsection{Donor-acceptor electronic transfer in rGO}

Using non-degenerate TA spectroscopy, we can excite molecular-like GQDs at high energies and probe the electron transfer rate to graphene at lower energy states. Figure 6a shows the normalized TA relaxation at 2.5 eV pump 1.2 eV probe near time zero, which shows a clear delayed rise in the most oxidized samples.  Conversely, the most reduced samples show a rise limited by the laser cross-correlation. This  delayed rising kinetic edge is indicative of an acceptor-donor electron relationship.     Charge transfer has been documented in GO, where photoexcited charges on a different molecular species are transferred to GO. \cite{Wang2010a,Zheng2015,Li2018} Figure 6b illustrates charge transfer between molecular GQDs and larger graphene-like regions. When the pump moved to longer energies (e.g. 1.8 eV in Fig. 1c), the delayed rise is no longer seen because the population of GQD donors is too small relative to graphene.

Figure 6b depicts the charge transfer process that is responsible for the observed delayed rise. Carriers photoexcited in the confined GQD states are localized by the surrounding oxygen functional groups. In GO, the large density of oxygenated regions results in a weaker coupling between confined GQDs and graphene submetallic sub-lattice regions, leading to the observed delayed rise. In the photoreduced samples, carriers excited into $sp^2$ GQD states are now closer to extendend graphene regions, and so the delayed acceptor-donor electron transfer is not observed to lower energy states. 

Figure 6c gives a qualitative description of the structure and acceptor-donor electron transfer process in rGO. Our graphene oxide begins with $\sim$44$\%$ oxygen content, these oxygen functional groups interrupt the delocalized $\pi$-orbitals and prohibit hopping between carbon sites. Reduction removes oxygen, which decreases the mean distance from a confined GQD donor and graphene-like  $sp^2$ sublattice region.  Such changes to the effective percolation network of the  $sp^2$ sublattice have previously been shown to also increase GO carrier mobility and conductivity \cite{Mohan2015,Wang2018}. The longer dynamics in GO are caused by excited carriers being more isolated by larger oxygenated regions as shown in Fig. 6c, which limit possible relaxation pathways. In rGO, some of the oxygen has been removed, recovering large-area graphene-like domains which decay more quickly than pristine graphene.

\section{Conclusions}

The highly variable  composition of  the quasi-amorphous  GO 2D lattice makes  a systematic  comparison against  monolayer  graphene a challenge.  To help overcome this challenge,  GO is suspended in a polymeric network scaffold where five successive photoreductions (rGO$_{1-5}$) were possible without any evidence of inter-layer aggregation.  Ultimately, this yielded optical quality rGO films with an absorption lineshape that fits to ml-graphene Fano resonance lineshape parameters. Likewise this step-wise photoreduction accelerates the hot electron relaxation kinetics monotonically over each of the variable probe energy windows studied from 1.2 to 2.5 eV.  At intermediate photoreduction times or rGO$_{2-3}$, Fig. 4 shows that a hot electron cooling model of disorder-assisted supercollision matches the $\tau_{SC}=$3.1 ps hot electron cooling of monolayer graphene. Figure 4b shows the recovery of ultrafast hot electron  relaxation rates similar to monolayer-graphene in moderately reduced samples(rGO$_{1-3}$ ), suggesting a largely uninterrupted $sp^2$ bonded network analogous to graphene. 

Under extreme photoreduction or using UV-Vis optical excitation, the optical properties of rGO begin to deviate strongly from graphene.  Owing to increasing local disorder and broken lattice symmetry, extreme photothermal reduction yields hot electron cooling rates that are faster than pristine graphene. Subsequent photoreduction accelerates the extracted hot electron cooling rate 10-12x, revealing how photodamage induces local disorder to mediate faster hot electron cooling.   On longer, $>$50 ps timescales, rGO also exhibits a slower decay response than graphene owing to many isolated graphene quantum dot (GQD) regions and oxygenated edge trap states which serve to delay the ground state recovery.  Using probe energies in the visible wavelength range at 1.8 eV, Figs. 1c and 4 shows that photothermal reduction does not recover pristine graphene properties, as evidenced by the slower decay kinetics of all rGO samples relative to graphene. The prevalence of isolated GQDs regions and oxygenated-edge trap states each create further bottlenecks of electronic relaxation that slow the effective relaxation.  Fortunately, we find these long lifetimes of rGO are no longer oberved below 1.3 eV optical excitations, as there are no discernible GQD sub-lattice states large enough to creae a resonance at these energies.  Collectively, these results show many of the desirable optoelectronics properties of 2D graphene can be replicated using selectively reduced graphene oxide suspended in a 3D bulk polymeric network. This study lends itself to large-scale processing of rGO thin films and applications in high-speed optoelectronics and photonic switching applications.


\begin{acknowledgments}
This material is based upon work supported by the Office of the Under Secretary of Defense for Research and Engineering under award number FA9550-22-1-0276, and the DEVCOM Army Research Laboratory award number W56HZV-16-C-0147. 
\end{acknowledgments}

\textbf{Supplementary Materials}: Details on sample characteristics, data modeling methods, and further absorption and  PL spectral data show similar graphene-like propertis out to  the mid-IR regions as far as 0.5 eV.  

\textbf{Data Availability Statement}:The data that support the findings of this study are available from the corresponding author upon reasonable request. 

\bibliography{rGO_BIBv1}

\end{document}